\documentclass[letterpaper, 10pt, conference]{ieeeconf} 
\IEEEoverridecommandlockouts 
\usepackage{cite}
\usepackage{setspace, epsfig, psfrag, amssymb, amsmath, color}
\usepackage{epstopdf,algorithm, algcompatible, lipsum, graphicx}

\definecolor{m1}{cmyk}{0, 0.99, 0.4429, 0.3412} 
\definecolor{m2}{cmyk}{0, 0.61, 0.4429, 0.2412} 
\definecolor{m3}{cmyk}{.83, 0, .90, .35} 
\definecolor{m4}{cmyk}{.83, 0, .40, .11} 
\definecolor{m4}{cmyk}{.83, 0, .40, .11} 
\definecolor{m5}{cmyk}{0,0.5,1,0} 
\def\bf{\normalfont\bfseries}
\def\bf{\normalfont\bfseries} 

\newfont{\smalll}{cmr8}
\def\IR{\mathbb{R}}

\def\span{\mathrm{span}}

\def\IC{\hbox{C\hskip-
.5em\raise.5ex\hbox{$\scriptscriptstyle\mid$}}\ }
\def\Ic{\hbox{\smalll C\hskip-
.5em\raise.3ex\hbox{$\scriptscriptstyle\mid$}}\ }

\def\T={\buildrel {\scriptscriptstyle\triangle} \over =}

\def\sqr#1#2{{\vcenter{\vbox{\hrule height.#2pt\hbox{\vrule
width.#2pt height#1pt \kern#1pt\vrule width.#2pt}\hrule
height.#2pt}}}}

\def\block-diag{\mathop{\rm block{\scriptstyle -}diag}}

\def\pmbb#1{\setbox0=\hbox{#1}\raise 0.5ex\box0}

\newcommand{\bequ}{\begin{eqnarray}}
\newcommand{\eequ}{\end{eqnarray}}
\newcommand{\mT}{^\mathrm{T}}

\newcommand{\rom}{\mathrm}

\newcommand {\beq}      {\begin{equation}}
\newcommand {\eeq}      {\end{equation}}

\def\IR{{\mathbb R}}
\def\IC{{\mathbb C}}

\renewcommand\theenumi{\roman{enumi}}
\renewcommand\theenumii{\alph{enumii}}
\renewcommand\theenumiii{\arabic{enumiii}}
\renewcommand\theenumiv{\Alph{enumiv}}

\def\sc{s_{\rm c}}

\def\sc{s_{\rm c}}

\definecolor{tBlue}{RGB}{25,100,250}

\definecolor{tRed}{RGB}{250,5,40}

\begin{document}
	{
		\title{{
				{\bf Regulating Stability Margins in Symbiotic Control:\\ A Low-Pass Filter Approach$^\star$}}}
	} 
	
	\author{Emre Yildirim$^\dagger$, Tansel Yucelen$^\dagger$, and John T. Hrynuk$^\ddagger$
		\thanks{$^\star$This research was supported by the United States Army Research Laboratory under the Grant {\tt{W911NF-23-S-0001.}}}
		\thanks{$^\dagger$Emre Yildirim and Tansel Yucelen are with the Laboratory for Autonomy, Control, Information, and Systems (LACIS, {\tt\footnotesize http:// lacis.eng.usf.edu/}) and the Department of Mechanical Engineering at the University of South Florida, Tampa, FL 33620, USA (emails: {\tt\footnotesize emreyildirim@usf.edu, \tt\footnotesize yucelen@usf.edu}).}
		\thanks{$^\ddagger$John T. Hrynuk is with the United States Army Research Laboratory, Adelphi, MD 20783, USA (email:{\tt\footnotesize john.t.hrynuk.civ@army.mil}).}
	}
	
	\maketitle
	
	
	
	\begin{abstract}
		
		Symbiotic control synergistically integrates fixed-gain control and adaptive learning architectures to mitigate system uncertainties more predictably than adaptive learning alone and without requiring prior knowledge of uncertainty bounds as compared to fixed-gain control alone. 
		Specifically, increasing the fixed-gain control parameter achieves a desired level of closed-loop system performance while the adaptive law simultaneously learns and suppresses the system uncertainties. 
		However, stability margins can be reduced when this parameter is large and this paper aims to address this practical challenge. 
		To this end, we propose a new fixed-gain control architecture predicated on a low-pass filter approach to regulate stability margins in the symbiotic control framework. 
		In addition to the presented system-theoretical results focusing on the stability of the closed-loop system, we provide two illustrative numerical examples to demonstrate how the low-pass filter parameters are chosen for the stability margin regulation problem without significantly compromising the closed-loop system performance. 
		
	\end{abstract}  
	
	
	\section{Introduction}\label{introduction}
	
	Fixed-gain control (e.g., robust control and sliding mode control) and adaptive learning (e.g., adaptive control and reinforcement learning) are two well-established architectures in control theory for mitigating the adverse effects of system uncertainties resulting from exogenous disturbances, parameter variations, and unmodeled dynamics. 
	Specifically, fixed-gain control architectures rely on the knowledge of uncertainty bounds (e.g., see [\citen{yedavalli2014robust}, Chapter 2] and [\citen{kurtoglu2022distributedsgn}, Remark 2]) and they are generally tuned to handle a worst-case scenario that may not happen in practice. 
	While they are conservative, these architectures result in predictable closed-loop system performance as the gains of the resulting control law are fixed. 
	On the other hand, adaptive learning architectures require minimal or no prior knowledge of such bounds and they are tuned online by a parameter adjustment mechanism for the purpose of achieving a desired level of closed-loop system performance. 
	However, due to their inherently nonlinear nature, these architectures can lead to less predictable closed-loop system performance, particularly during the learning phase (e.g., see [\citen{yucelen2012low}] and [\citen{gibson2013adaptive}]). 
	
	Symbiotic control builds on the strengths of both fixed-gain control and adaptive learning architectures by synergistically integrating them to mitigate system uncertainties in a more predictable manner than adaptive learning alone, while also eliminating the need for any prior knowledge of uncertainty bounds typically required by fixed-gain control. 
	In particular, symbiotic control is recently proposed in [\citen{yucelen2024symbioticARXIV}] for dynamical systems with parametric and nonparametric uncertainties. 
	While introduced in [\citen{yucelen2024symbioticARXIV}], the foundation of this framework is laid by earlier work including [\citen{yucelen2013new}], [\citen{gruenwald2015transient}], and [\citen{gruenwald2017direct}]. 
	To improve closed-loop system performance for dynamical systems with parametric uncertainties, [\citen{yucelen2013new}] focuses on altering the trajectories of the reference signal, [\citen{gruenwald2015transient}] introduces artificial basis functions, and [\citen{gruenwald2017direct}] presents a gradient descent optimization framework. 
	Of these, the findings reported in [\citen{gruenwald2017direct}] align most closely with the recent results in [\citen{yucelen2024symbioticARXIV}], although [\citen{gruenwald2017direct}] assumes some knowledge of uncertainty bounds to guarantee closed-loop stability (see [\citen{gruenwald2017direct}, (34)]). 
	This assumption is not only removed in [\citen{yucelen2024symbioticARXIV}] but the results also generalize to dynamical systems with nonparametric uncertainties.
	
	\begin{figure}[h!] \vspace{-0.25cm}
		\center \includegraphics[width=0.275\textwidth]{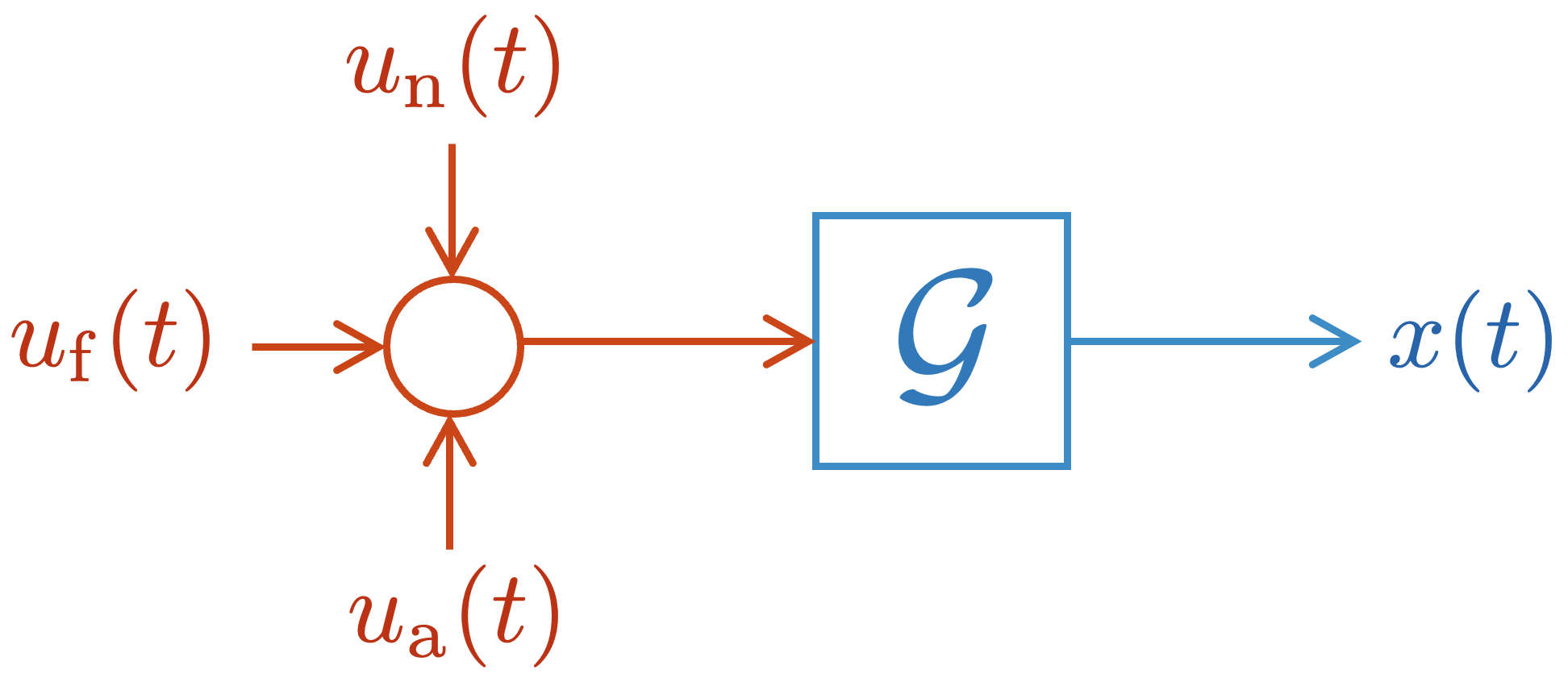} \vspace{-0.25cm}
		\caption{Key input signals of the symbiotic control framework for driving an uncertain dynamical system $\mathcal{G}$ with $u_\rom{n}(t)$ denoting the nominal control, $u_\rom{f}(t)$ denoting the fixed-gain control, and $u_\rom{a}(t)$ denoting the adaptive control, where $x(t)$ is the state.} \vspace{-0.25cm}
		\label{symbiotic:control:figure} 
	\end{figure}
	
	Figure \ref{symbiotic:control:figure} shows the key input signals of the symbiotic control framework. 
	Specifically, the nominal control $u_\rom{n}(t)$ is designed based on the known portion of the uncertain dynamical system, or it can be an existing legacy control structure. 
	It ensures a baseline closed-loop system performance under nominal operating conditions (i.e., no system uncertainties). 
	To mitigate the adverse effects of system uncertainties, the fixed-gain control $u_\rom{f}(t)$ and the adaptive control $u_\rom{a}(t)$ augment the nominal control. 
	In the presence of system uncertainties and the absence of adaptive control, [\citen{yucelen2024symbioticARXIV}, Proposition 1] utilizes singular perturbation theory to show that the uncertain dynamical system achieves its baseline closed-loop system performance when the fixed-gain control parameter is sufficiently large. 
	Yet, since it is not possible to know in practice how large this parameter needs to be, it is shown in [\citen{yucelen2024symbioticARXIV}, Theorems 1--3] that increasing the fixed-gain control parameter achieves a desired level of closed-loop system performance while the adaptive law simultaneously learns and suppresses the remaining system uncertainties of parametric and nonparametric nature without requiring any knowledge of their bounds. 
	
	From a practical standpoint, these results imply that the fixed-gain control parameter should be initially set to a candidate value and then increased until the adverse effects of system uncertainties are sufficiently mitigated. 
	In addition, the adaptive control parameters should be chosen smaller relative to the fixed-gain control parameter to ensure that adaptive control ultimately mitigates the remaining system uncertainties. 
	It is also reported in [\citen{yucelen2024symbioticARXIV}] that fixed-gain control can achieve a desired level of closed-loop system behavior even with an insufficient number of neurons or when high leakage term parameters are utilized within the adaptive control design. 
	This highlights the importance of fixed-gain control in the symbiotic control framework as it serves as the foundation for ensuring resilient and predictable closed-loop system performance. 
	
	While a large fixed-gain control parameter is important for the symbiotic control framework, it can lead to reduced stability margins and the purpose of this paper is to address that practical challenge. 
	In particular, we propose a new fixed-gain control architecture predicated on a low-pass filter approach\footnote{While [\citen{yucelen2013new}, Theorem 7.1] and [\citen{gruenwald2015transient}, Corollary 2] consider different low-pass filtering approaches, they focus on reducing high-frequency oscillations and do not discuss the regulation of stability margins problem considered here.} to regulate stability margins within the symbiotic control framework. 
	The results presented here focus on mitigating the effects of exogenous disturbances, which allows for a linear setting that facilitates direct analysis of stability margins. 
	Although this differs from the nonlinear setting in [\citen{yucelen2024symbioticARXIV}], where the focus is on mitigating unknown parameter variations, the proposed fixed-gain control architecture of this paper can be directly used as-is within the framework of [\citen{yucelen2024symbioticARXIV}], and therefore, there is no loss of generality in considering the exogenous disturbance problem\footnote{This is because the main component of the fixed-gain control law, which is responsible for mitigating the adverse effects of system uncertainties, already has a linear form (see [\citen{yucelen2024symbioticARXIV}, (7)]. As a consequence, one can directly replace it with the proposed fixed-gain control structure of this paper.}. 
	In addition to the system-theoretical results on the stability of the closed-loop system, two illustrative numerical examples are further provided to demonstrate how the low-pass filter parameters are selected for the stability margin regulation problem without significantly compromising the closed-loop system performance.
	
	
	\section{Mathematical Preliminaries}\label{problem_formulation}
	
	We begin with the notation used in this paper. 
	Specifically, we respectively use $\IR$, $\IR^n$, and $\IR^{n \times m}$ for sets of real numbers, real vectors, and real matrices; $\IR_{+}$ and $\IR^{n \times n}_{+}$ for the sets of positive real numbers and positive-definite real matrices; and ``$\triangleq$'' for the equality by definition. 
	In addition, $(\cdot)^{-1}$ denotes the inverse, $(\cdot)\mT$ denotes the transpose, $\left\|\cdot\right\|_{2}$ denotes the vector Euclidean norm or the matrix-induced 2-norm, 
	and $\underline{\lambda}\left(A\right)$ and $\overline{\lambda}\left(A\right)$ respectively denote the minimum and the maximum eigenvalues of the real matrix $A\in\IR^{n\times n}$. 
	
	Next, we present a concise overview on how the results documented in [\citen{yucelen2024symbioticARXIV}] apply to the exogenous disturbance problem. 
	To this end, consider the dynamical system represented in the state-space form given by
	\bequ 
	\ \ \ \dot{x}(t)=A x(t) + B \bigl(u(t) +d(t)\bigl), \quad x(0)=x_0, \label{2.1}
	\eequ 
	where $x(t)\in\IR^n$ denotes the state vector, 
	$u(t)\in \IR^m$ denotes the control input, 
	$d(t)\in \IR^m$ denotes an unknown bounded exogenous disturbance with a bounded time rate of change\footnote{The exogenous disturbance considered here is matched, which occurs in physical systems when external forces are applied through the same mechanisms as the control inputs. 
		A considerable number of physical systems are subject to matched exogenous disturbances such as external forces acting on robotic arms through the same joints as the control actuators and winds affecting the control surfaces of an aircraft [\citen{yucelenMRAC}, Section 2].}, and $A\in \IR^{n\times n}$ and $B\in \IR^{n\times m}$ respectively denote the known system and full column rank control matrices with the pair $(A,B)$ being stabilizable. 
	
	As shown in Figure \ref{symbiotic:control:figure}, the control input consists of three key input signals 
	\bequ 
	u(t) = u_\rom{n}(t) +u_\rom{f}(t) +u_\rom{a}(t), \label{control_input}
	\eequ 
	where these inputs are discussed next. 
	In particular, the nominal control $u_\rom{n}(t)\in\IR^m$ has the form\footnote{We here employ a static nominal control given by (\ref{2.2}). If desired, a dynamic nominal control can be utilized instead, with only slight modifications required to the results presented in this paper (see [\citen{yucelenMRAC}, Section 3.2] for an example).} given by  
	\bequ 
	u_\rom{n}(t) = -K_1 x(t)+K_2 r(t), \label{2.2}
	\eequ 
	where $K_1\in\IR^{m \times n}$ and $K_2\in\IR^{m \times p}$ respectively denote a feedback gain matrix with $A-BK_1$ being Hurwitz and a feedforward gain matrix. 
	In (\ref{2.2}), $r(t)\in\IR^p$ denotes a given uniformly continuous and bounded reference signal. 
	As discussed in the third paragraph of Section \ref{introduction}, the nominal control ensures a baseline closed-loop system performance under nominal operating conditions (i.e., in the absence of exogenous disturbance), where this ideal performance is captured by the reference model given by 
	\bequ
	\ \ \ \ \ \dot{x}_\rom{n}(t)=A_\rom{n}x_\rom{n}(t)+B_\rom{n}r(t), \quad x_\rom{n}(0)=x_{\rom{n}{0}}, \label{reference_model}
	\eequ 
	with $A_\rom{n}\triangleq A-BK_1$ and $B_\rom{n}\triangleq B K_2$. 
	
	To mitigate the adverse effects of the exogenous disturbance, the fixed-gain control $u_\rom{f}(t)\in\IR^m$ has the form
	\bequ
	u_\rom{f}(t)&\hspace{-0.34cm}=\hspace{-0.34cm}&-\alpha B_\rom{i}(x(t)\hspace{-0.05cm}-\hspace{-0.05cm}x_0)\hspace{-0.08cm}+\hspace{-0.08cm}\alpha B_\rom{i}\hspace{-0.10cm}\int_0^t \hspace{-0.15cm}\bigl(A_\rom{n}x(s)\hspace{-0.08cm}+\hspace{-0.08cm}B_\rom{n}r(s)\bigl)\rom{d}s, \ \ \  \label{fixed_gain_control_equiv}
	\eequ 
	where $\alpha\in \IR_+$ denotes the fixed-gain control parameter and $B_\rom{i}\triangleq (B\mT B)^{-1}B\mT$. 
	Note that the inverse of $B\mT B$ exists since $B$ has full column rank. 
	In addition, the adaptive control $u_\rom{a}(t)\in\IR^m$ has the form
	\bequ 
	u_\rom{a}(t) = -\hat{d}(t)
	\eequ 
	with $\hat{d}(t)$ being the learning estimate of $d(t)$ that satisfies the parameter adjustment mechanism given by 
	\bequ
	\dot{\hat{d}}(t) =  \beta_3^{-1}\big(\beta_1 e\mT(t)PB\hspace{-0.05cm}-\hspace{-0.05cm}\alpha \beta_2 u_\rom{f}\mT(t)\big)\mT\hspace{-0.10cm}-\hspace{-0.05cm}\mu_1 \hat{d}(t), \nonumber \\  \hat{d}(0)=\hat{d}_{0}. \label{uncer_estimate}
	\eequ
	In (\ref{uncer_estimate}), $e(t)\triangleq x(t)-x_\rom{n}(t)\in \IR^n$ denotes the error signal; $\beta_1\in\IR_+$, $\beta_2\in\IR_+$, and $\beta_3\in\IR_+$ denote learning parameters;  $\mu_1\in \IR_+$ denotes the leakage parameter; and $P\in \IR_+^{n\times n}$ denotes the unique  solution to the Lyapunov equation given by 
	\vspace{-0.05cm}
	\bequ 
	0=A\mT_\rom{n}P+PA_\rom{n}+R \label{L:E}
	\eequ 
	with $R\in\IR_+^{n\times n}$. 
	
	
	Defining the disturbance error as $\tilde{d}(t)\triangleq\hat{d}(t)-d(t)\in\IR^m$, one can now rewrite (\ref{2.1}) as
	\bequ 
	\dot{x}(t)\hspace{-0.05cm}=\hspace{-0.05cm}A_\rom{n}x(t)+B_\rom{n}r(t)+B\bigl(u_\rom{f}(t)-\tilde{d}(t)\bigl). \label{compact_dynamics}
	\eequ 
	The following facts are now immediate: 
	\begin{itemize}
		\item If $\alpha$ in (\ref{fixed_gain_control_equiv}) is sufficiently large, then the solution to (\ref{compact_dynamics}) approximately behaves as the solution to the reference model (\ref{reference_model}). 
		\item If the exogenous disturbance is constant and the leakage parameter is zero, then all closed-loop signals are bounded and $\lim_{t\rightarrow\infty} e(t)=0$. 
		\item If the exogenous disturbance is time-varying and the leakage parameter is positive, then all closed-loop signals are bounded.  
	\end{itemize}
	
	
	\textbf{Remark 1.} Note that the first fact is from [\citen{yucelen2024symbioticARXIV}, Theorem 1] and the second fact is from [\citen{yucelen2024symbioticARXIV}, Theorem 2]. 
	Note also that the third fact follows similar steps as in the proof of [\citen{yucelen2024symbioticARXIV}, Theorem 3], where the aforementioned bound can be rigorously quantified in a form similar to [\citen{yucelen2024symbioticARXIV}, (43)]. 
	Regardless of whether the exogenous disturbance is constant or time-varying and whether the leakage term is zero or positive, the first fact shows that the closed-loop system behavior becomes more predictable as $\alpha$ increases.
	
	
	\textbf{Remark 2.} While the solution to (\ref{compact_dynamics}) approximately behaves as the solution to the reference model (\ref{reference_model}) as $\alpha$ increases, it can lead to reduced stability margins as also discussed in the last paragraph of Section \ref{introduction}. 
	The next section introduces a new fixed-gain control architecture predicated on a low-pass filter approach to address this problem, where it also system-theoretically shows the stability of the resulting closed-loop system.
	Section \ref{numerical_examples} then illustrates how this low-pass filter is important to regulate stability margins in the symbiotic control framework. 
	
	
	
	
	
	\section{A Low-Pass Filter Approach to\\Symbiotic Control}\label{system_theory}
	
	Consider the symbiotic control framework presented in the previous section for the mitigation of the exogenous disturbance problem. 
	However, instead of (\ref{fixed_gain_control_equiv}), consider the new fixed-gain control architecture given by
	\bequ
	u_\rom{f}(t)&\hspace{-0.9cm}=\hspace{-0.9cm}&-\alpha B_\rom{i}(x(t)\hspace{-0.05cm}-\hspace{-0.05cm}x_0)\hspace{-0.08cm} +\alpha B_\rom{i}\hspace{-0.10cm}\int_0^t \hspace{-0.15cm}\bigl(A_\rom{n}x(s)\hspace{-0.08cm}+\hspace{-0.08cm}B_\rom{n}r(s)\bigl)\rom{d}s \nonumber \\ && -\alpha \epsilon_1\hspace{-0.1cm}\int_0^t \hspace{-0.15cm}\bigl(u_\rom{f}(s)-u_\rom{fl}(s)\bigl)\rom{d}s,  \label{fixed_gain_control_equiv2}
	\eequ 
	where $\epsilon_1\in\IR_+$ is an auxiliary fixed-gain parameter and $u_\rom{fl}(t)$ denotes a low-pass filter satisfying
	\bequ
	\dot{u}_\rom{fl}(t) = - \epsilon_2\big(u_\rom{fl}(t)-u_\rom{f}(t)\big)-\mu_2 u_\rom{fl}(t), \  u_\rom{fl}(0)=u_{\rom{fl}{0}},\label{fixed_gain_control_ufl} 
	\eequ
	with $\epsilon_2\in\IR_+$ being the low-pass filter parameter and $\mu_2\in\IR_+$ being the leakage parameter. 
	Note that the time constant and the gain of this low-pass filter are respectively given by $\tau \triangleq 1/(\epsilon_2 + \mu_2)$ and $\mathcal{K} \triangleq \epsilon_2/(\epsilon_2 + \mu_2)$. 
	Note also that the leakage parameter $\mu_2$ should be chosen sufficiently small such that the low-pass filter gain $\mathcal{K}$ stays close to unity\footnote{Otherwise, the low-pass filter gain $\mathcal{K}$ can get smaller and $u_\mathrm{fl}(t)$ is forced to remain near its zero equilibrium.}. 
	We are now ready to present a key lemma. 
	
	\textbf{Lemma 1.} The new fixed-gain control architecture given by 
	(\ref{fixed_gain_control_equiv2}) is equivalent to
	\vspace{0cm}
	\bequ 
	\dot{u}_\rom{f}(t) = -\alpha\big(u_\rom{f}(t)-\tilde{d}(t)\big)-\alpha \epsilon_1\big(u_\rom{f}(t)-u_\rom{fl}(t)\big), \nonumber \\  u_\rom{f}(0)=u_{\rom{f}{0}},\label{fixed_gain_control_uf} 
	\eequ 
	
	\textit{Proof.} One can write 
	\bequ
	u_\rom{f}(t)-\tilde{d}(t) = B_i\big(\dot{x}(t)-A_\rom{n}x(t)-B_\rom{n}r(t)\big) \hspace{-0.25cm}\label{Equiv_1}
	\eequ
	by multiplying both sides of (\ref{compact_dynamics}) by $B_i$. 
	Now, using (\ref{Equiv_1}) in (\ref{fixed_gain_control_uf}) yields 
	\bequ
	\dot{u}_\rom{f}(t)&\hspace{-0.35cm}=\hspace{-0.35cm}& -\alpha B_i\big(\dot{x}(t)-A_\rom{n}x(t)-B_\rom{n}r(t)\big)\nonumber 
	\\&&- \alpha \epsilon_1 \big( u_\rom{f}(t)-u_\rom{fl}(t)\big), \ \ u_\rom{f}(0)=u_{\rom{f}{0}}.\label{Equiv_2}
	\eequ
	Finally, taking the integral of (\ref{Equiv_2}) gives (\ref{fixed_gain_control_equiv2}). 
	The proof is now complete. \hfill $\blacksquare$
	
	\textbf{Remark 3.} While (\ref{fixed_gain_control_equiv2}) is equivalent to (\ref{fixed_gain_control_uf}), the latter is not implementable in practical applications since $d(t)$ is unknown and it is only needed for analyzing the stability of the closed-loop system presented later in this section. 
	
	\textbf{Remark 4.} The following observations about the new fixed-gain control architecture given by (\ref{fixed_gain_control_equiv2}) (or equivalently (\ref{fixed_gain_control_uf})) and (\ref{fixed_gain_control_ufl}) are now necessary: 
	\begin{itemize}
		\item  The first term on the right side of (\ref{fixed_gain_control_uf}) forces $u_\mathrm{f}(t)$ to behave like $\tilde{d}(t)$ as $\alpha$ increases, and $\epsilon_1$ decreases and/or $\epsilon_2$ increases\footnote{The new fixed-gain control architecture behaves as its original version given by (\ref{fixed_gain_control_equiv}) as $\epsilon_1$ decreases and/or $\epsilon_2$ increases.}. 
		This is important to mitigate the effect of the disturbance error $\tilde{d}(t)$, which denotes the mismatch between the exogenous disturbance $d(t)$ and its learning estimate $\hat{d}(t)$. 
		\item The second term on the right side of (\ref{fixed_gain_control_uf}) forces $u_\mathrm{f}(t)$ to behave like its low-pass filter version $u_\mathrm{fl}(t)$ as the product of $\alpha$ and $\epsilon_1$ increases. 
		Forcing $u_\mathrm{f}(t)$ to approach $u_\mathrm{fl}(t)$ is important to limit the aggressive behavior of the fixed-gain control law, which in turn helps one to regulate stability margins of the symbiotic control framework (see Section \ref{numerical_examples}). 
	\end{itemize}
	The above two points highlight the tradeoff between closed-loop system performance and stability margins through the adjustment of the parameters $\alpha$, $\epsilon_1$, and $\epsilon_2$.
	
	For the next result on the stability of the closed-loop system, we write the error dynamics as 
	\bequ
	\dot{e}(t) = A_\rom{n}e(t)+B u_\rom{f}(t)-B\tilde{d}(t), \quad e(0)=e_0, \label{error_dynamics}
	\eequ
	using (\ref{reference_model}) and (\ref{compact_dynamics}). 
	In addition, we introduce the Lyapunov function candidate as
	\bequ 
	\mathcal{V}(e(t),u_\rom{f}(t),\tilde{d}(t),u_\rom{fl}(t)) & \hspace{-0.25cm}= \hspace{-0.25cm}&  \delta \mT(t) M_1\delta(t)+\beta_3 \tilde{d}\hspace{0.05cm}\mT\hspace{-0.05cm}(t)\tilde{d}(t) \nonumber \\&&+ \beta_4u_\rom{fl}\mT(t)u_\rom{fl}(t), \label{Lyap_func}
	\eequ
	with $\beta_1\in\IR_+$, $\beta_2\in\IR_+$, $\beta_3\in\IR_+$, $\beta_4\triangleq \epsilon_2^{-1}\alpha \beta_2\epsilon_1\in\IR_+$, $\delta(t) \hspace{-0.1cm} \triangleq \hspace{-0.1cm} [e\mT(t),u\mT_\rom{f}(t)]\mT$, and $M_1 \triangleq \rom{diag}([\beta_1 P, \beta_2 I])$. 
	Observe that $\mathcal{V}(0,0,0,0)=0$, $\mathcal{V}(e(t),u_\rom{f}(t),\tilde{d}(t),u_\rom{fl}(t))\in\IR_+$, and $\mathcal{V}(e(t),u_\rom{f}(t),\tilde{d}(t),u_\rom{fl}(t))$ is radially unbounded. 
	Finally, let $\mathcal{V}_0 \hspace{-0.05cm}\triangleq  \mathcal{V}(e(0),u_\rom{f}(0),\tilde{d}(0),u_\rom{fl}(0))$, $b^\star\triangleq \rom{min}\{l_1,l_2,l_3\}$, $l_1\hspace{-0.05cm}\triangleq\underline{\lambda}(M_3){\overline{\lambda}}^{-1}(M_1)\in\IR_+$, $l_2\triangleq 2\mu_1-\mu_1d_1-d_2\in\IR_+$, $l_3\triangleq \underline{\lambda}(M_2)\beta_4^{-1}\in\IR_+$, $l^\star\triangleq \beta_3\mu_1 d_1^{-1}\bar{d}\hspace{0.04cm}^2+\beta_3 d_2^{-1}\bar{\dot{d}}^{\hspace{0.05cm}2}\in\IR_+$, $d_1\in\IR_+$, $d_2\in\IR_+$, $||d(t)||_2\leq \bar{d}$, $||\dot{d}(t)||_2\leq \bar{\dot{d}}$, and 
	\bequ
	\hspace{-0.55cm}M_2\hspace{-0.30cm}&\triangleq&\hspace{-0.30cm}\begin{bmatrix}2\alpha\beta_2\epsilon_1 & -2\beta_4\epsilon_2 \\ -2\beta_4\epsilon_2 & 2\beta_4\epsilon_2+ 2\beta_4\mu_2\end{bmatrix}\hspace{-0.125cm}\in\hspace{-0.05cm}\IR_+^{2 \times 2}, \\
	\hspace{-0.55cm}M_3\hspace{-0.30cm}&\triangleq&\hspace{-0.30cm}\begin{bmatrix}\beta_1 R & \hspace{-0.2cm}-\beta_1PB\\ -\beta_1B\mT P & \hspace{-0.2cm}(2\beta_2\alpha+\underline{\lambda}(M_2))I\end{bmatrix}\hspace{-0.125cm}\in\hspace{-0.05cm}\IR_+^{(n+m) \times (n+m)}\hspace{-0.05cm}. \
	\eequ
	Note that the positiveness of $l_2$ is ensured since $d_1$ and $d_2$ are arbitrary positive constants, 
	the positive-definiteness of $M_2$ is ensured by [\citen{Lewis_book}, Lemma 3.3] based on the above selection for $\beta_4$, 
	and the positive-definiteness of $M_3$ can be ensured since a sufficiently large $\beta_2 \alpha$ can always be chosen. 
	We are now ready to present the following result. 
	
	\textbf{Theorem 1.} Consider the uncertain dynamical system given in (\ref{2.1}), the dynamics of the reference model given in (\ref{reference_model}), and the feedback control law (\ref{control_input}) along with (\ref{2.2}), (\ref{uncer_estimate}), (\ref{fixed_gain_control_uf}) and (\ref{fixed_gain_control_ufl}). 
	Then, the solution \big($e(t)$, $u_\rom{f}(t)$, $\tilde{d}(t)$, $u_\rom{fl}(t)$\big) of the closed-loop system for all \big($e_0$, $u_{\rom{f}0}$, $\tilde{d}_0$, $u_{\rom{fl}0}$\big) is bounded\footnote{Recall to choose $\beta_2$ of (\ref{uncer_estimate}) and $\alpha$ of (\ref{fixed_gain_control_equiv2}) properly to guarantee that the product $\beta_2\alpha$ yields $M_3\in\IR_+^{2 \times 2}$.} according to 
	\bequ
	\mathcal{V}(\cdot) &\hspace{-0.25cm}\leq\hspace{-0.25cm} & \mathcal{V}_0 \rom{exp}(-b^\star t)+ \frac{l^\star}{b^\star}. \label{Bound} 
	\eequ
	
	
	
	\textit{Proof.} The time derivative of the Lyapunov function candidate given by (\ref{Lyap_func}) satisfies
	\vspace{0cm}
	\bequ 
	\dot{\mathcal{V}}(\cdot) \hspace{-0.25cm}&=& \hspace{-0.25cm}
	2 \beta_1 e\mT(t) P \bigl[A_\rom{n}e(t)\hspace{-0.05cm}+\hspace{-0.05cm}Bu_\rom{f}(t) 
	-B\tilde{d}(t)\bigl] \nonumber\\
	&&\hspace{-0.25cm}+2\beta_2 u_\rom{f}\mT(t)\big[\hspace{-0.07cm}-\hspace{-0.07cm}\alpha \bigl(u_\rom{f}(t)\hspace{-0.05cm}-\hspace{-0.05cm}\tilde{d}(t)\bigl)-\alpha \epsilon_1\hspace{-0.05cm}\big(u_\rom{f}(t)-u_\rom{fl}(t)\big)\big] \nonumber\\
	&& \hspace{-0.25cm}+ 2 \beta_3 \tilde{d}\hspace{0.05cm}\mT\hspace{-0.05cm}(t)\big(\dot{\hat{d}}(t)\hspace{-0.03cm}-\hspace{-0.03cm}\dot{d}(t)\big)\hspace{-0.075cm}-\hspace{-0.075cm}2\beta_4\epsilon_2 u_\rom{fl}\mT(t)\big(u_\rom{fl}(t)\hspace{-0.075cm}-\hspace{-0.075cm}u_\rom{f}(t)\big)\nonumber\\
	&&\hspace{-0.25cm}-2\beta_4 \mu_2 u_\rom{fl}\mT(t)u_\rom{fl}(t). 
	\label{V_dot1}
	\eequ
	Using (\ref{uncer_estimate}) in (\ref{V_dot1}) now yields
	\bequ
	\dot{\mathcal{V}}(\cdot) &\hspace{-0.2cm}=\hspace{-0.1cm}&\hspace{-0.15cm}-\beta_1e\mT(t) R e(t) + 2 \beta_1 e\mT(t) P B u_\rom{f}(t)   \nonumber 
	\\ && -2\alpha \beta_2u_\rom{f}\mT(t)u_\rom{f}(t) -2 \alpha \beta_2\epsilon_1 u_\rom{f}\mT(t)u_\rom{f}(t) \nonumber 
	\\&& + 2 \alpha \beta_2 \epsilon_1 u_\rom{f}\mT(t)u_\rom{fl}(t)-2\beta_4\mu_2u_\rom{fl}\mT(t)u_\rom{fl}(t) \nonumber 
	\\&& -2 \beta_4 \epsilon_2 u_\rom{fl}\mT(t)u_\rom{fl}(t) + 2\beta_4\epsilon_2u_\rom{fl}\mT(t) u_\rom{f}(t) \nonumber 
	\\&& -2 \beta_3 \tilde{d}\hspace{0.05cm}\mT\hspace{-0.05cm}(t)\dot{d}(t)-2\beta_3\mu_1 \hat{d}\hspace{0.05cm}\mT\hspace{-0.05cm}(t)\tilde{d}(t)    .       \label{V_dot2}
	\eequ 
	
	Next, we apply Young's inequality to the following two sign-indefinite terms as
	\bequ
	-2\beta_3\mu_1d\mT(t)\tilde{d}(t)\leq 2\beta_3\mu_1(\frac{d_1}{2}||\tilde{d}(t)||_2^2+\frac{1}{2d_1}\bar{d}\hspace{0.05cm}^2), \label{Young1} \\
	-2\beta_3\tilde{d}\hspace{0.05cm}\mT(t)\dot{d}(t)\leq 2 \beta_3(\frac{d_2}{2}||\tilde{d}(t)||_2^2+\frac{1}{2d_2}\bar{\dot{d}}\hspace{0.05cm}^2). \label{Young2}
	\eequ
	Using now (\ref{Young1}) and (\ref{Young2}) in (\ref{V_dot2}), one obtains
	\bequ
	\dot{\mathcal{V}}(\cdot) &\hspace{-0.2cm}\leq \hspace{-0.1cm}&\hspace{-0.15cm}-\beta_1e\mT\hspace{-0.05cm}(t) Re(t) + 2 \beta_1 e\mT(t) P B u_\rom{f}(t)   \nonumber 
	\\ && -2\alpha \beta_2u_\rom{f}\mT(t)u_\rom{f}(t) -l_2\beta_3||\tilde{d}(t)||_2^2 \nonumber
	\\&& -\vartheta\mT(t) M_2\vartheta(t)+l^\star,       \label{V_dot3}
	\eequ
	with $\vartheta(t)\triangleq[u_\rom{f}\mT(t),u_\rom{fl}\mT(t)]\mT$, where (\ref{V_dot3}) can equivalently be rewritten as
	\vspace{0cm}
	\bequ
	\dot{\mathcal{V}}(\cdot) &\hspace{-0.2cm}\leq \hspace{-0.1cm}&\hspace{-0.15cm}-\delta\mT\hspace{-0.05cm}(t)M_3\delta(t) -l_2\beta_3||\tilde{d}(t)||_2^2 \nonumber
	\\&& -\underline{\lambda}(M_2)u_\rom{fl}\mT(t)u_\rom{fl}(t)+l^\star.       \label{V_dot4}
	\eequ
	
	Finally, consider $\delta\mT(t)M_1\delta(t)$ in (\ref{Lyap_func}) and $\delta\mT(t)M_3\delta(t)$ in (\ref{V_dot4}), where one can write
	\bequ
	-\delta\mT(t)M_3\delta(t)\leq -\frac{\underline{\lambda}(M_3)}{\overline{\lambda}(M_1)}\delta\mT(t) M_1 \delta(t). \label{upper_bound}
	\eequ
	An upper bound to (\ref{V_dot4}) can now be given as
	\bequ
	\dot{\mathcal{V}}(\cdot) &\hspace{-0.25cm}\leq \hspace{-0.25cm}& 
	\hspace{-0.05cm}-\hspace{-0.0005cm}l_1\delta\mT\hspace{-0.05cm}(t)M_1\delta(t) 
	\hspace{-0.05cm}-\hspace{-0.0005cm}l_2\beta_3||\tilde{d}(t)||_2^2
	\hspace{-0.05cm}-\hspace{-0.0005cm}l_3\beta_4 u_\rom{fl}\mT(t)u_\rom{fl}(t)
	\hspace{-0.05cm}+\hspace{-0.0005cm}l^\star \nonumber\\
	&\hspace{-0.25cm}\leq \hspace{-0.25cm}& -b^\star \mathcal{V}(\cdot)+l^\star
	\label{V_dot5}
	\eequ
	Applying the comparison principle to (\ref{V_dot5}), (\ref{Bound}) is now immediate. 
	The proof is now complete. \hfill $\blacksquare$
	
	In the next result, we show that the error signal and the fixed-gain control signal approach zero for the case when the exogenous disturbance is constant. 
	For this purpose, let 
	\bequ
	M_4\triangleq\begin{bmatrix}\beta_1 R & -\beta_1PB\\ -\beta_1B\mT P & 2\beta_2\alpha I\end{bmatrix}\in\IR_+^{(n+m) \times (n+m)}.
	\eequ
	Once again, the positive-definiteness of $M_4$ can be ensured since a sufficiently large  $\beta_2 \alpha$ can always be chosen.
	We are now ready to present the following result. 
	
	\textbf{Theorem 2.} Under the consideration in Theorem 1, 
	\bequ
	\lim_{t\rightarrow \infty} \bigl(e(t), \hspace{0.03cm}u_\rom{f}(t)\bigl)=\bigl(0, \hspace{0.03cm}0\bigl) \label{corollary:1}
	\eequ
	holds\footnote{Recall to choose $\beta_2$ of (\ref{uncer_estimate}) and $\alpha$ of (\ref{fixed_gain_control_equiv2}) properly to guarantee that the product $\beta_2\alpha$ yields $M_4\in\IR_+^{2 \times 2}$.} when the exogenous disturbances are constant and the leakage parameters (i.e., $\mu_1$ and $\mu_2$) are zero.
	
	
	\textit{Proof.} When $\bar{\dot{d}}=0$, $\mu_1=0$, and $\mu_2=0$, the time derivative of the Lyapunov function candidate (\ref{Lyap_func}) satisfies
	\bequ 
	\dot{\mathcal{V}}(\cdot) &\hspace{-0.6cm}=\hspace{-0.5cm}& 
	- \beta_1 e\mT\hspace{-0.05cm}(t) R e(t)\hspace{-0.05cm}+\hspace{-0.05cm}2\beta_1e\hspace{-0.05cm}\mT(t)PBu_\rom{f}(t)\hspace{-0.05cm}+\hspace{-0.05cm}2\beta_3\tilde{d}\hspace{0.05cm}\mT\hspace{-0.05cm}(t)\dot{\hat{d}}(t) \nonumber
	\\&&-2\beta_1 e\mT\hspace{-0.05cm}(t)PB\tilde{d}(t)-2\alpha\beta_2u_\rom{f}\mT(t)u_\rom{f}(t) \nonumber
	\\&& +2\alpha\beta_2u_\rom{f}\mT(t)\tilde{d}(t)-2\alpha\beta_2\epsilon_1||u_\rom{fl}(t)-u_\rom{f}(t)||_2^2.
	\label{Vcostant_dot1}
	\eequ
	Using (\ref{uncer_estimate}) in (\ref{Vcostant_dot1}) now yields
	\bequ 
	\dot{\mathcal{V}}(\cdot) &\hspace{-0.25cm}=\hspace{-0.25cm}& 
	-\delta\mT(t)M_4\delta(t) -2\alpha\beta_2\epsilon_1||u_\rom{fl}(t)-u_\rom{f}(t)||_2^2 \nonumber\\
	&\hspace{-0.25cm}\le \hspace{-0.25cm}&-\delta\mT(t)M_4\delta(t).
	\label{Vcostant_dot2}
	\eequ
	From the LaSalle-Yoshizawa theorem [\citen{haddad2008nonlinear}, Theorem 4.7], (\ref{corollary:1}) is immediate. 
	The proof is now complete. \hfill $\blacksquare$
	
	
	\section{Illustrative Numerical Examples}\label{numerical_examples}
	
	In this section, we present two illustrative numerical examples to show the efficacy of the new fixed-gain control architecture. 
	Specifically, we resort to the classical control theory tools in these examples to analyze the stability margins obtained from the loop transfer function (broken at the control input) and compare the standard fixed-gain control law given by (\ref{fixed_gain_control_equiv}) with the new fixed-gain control law given in (\ref{fixed_gain_control_equiv2}) and (\ref{fixed_gain_control_ufl}). 
	In both examples, we consider $A=\begin{bmatrix}0 & 1 \\ 0 & 0 \end{bmatrix}$ and $B=\begin{bmatrix}0 \\ 1 \end{bmatrix}$ for (\ref{2.1}). 
	We also choose $K_1=[0.16, 0.57]$, $K_2=0.16$, and $R=I$. 
	In addition, the learning parameters are selected as $\beta_1\hspace{-0.05cm}=\hspace{-0.05cm}0.1$, $\beta_2=3$, $\beta_3=1$, and $\beta_4=0.1$, which guarantee that $M_3$ and $M_4$ are positive-definite. 
	The leakage parameters $\mu_1$ and $\mu_2$ are further set to zero as we consider a constant disturbance\footnote{Recall that a constant or time-varying disturbance does not affect the computation of stability margins.} here, which is given by $d(t)=10$. 
	Finally, the reference signal $r(t)$ is given as a filtered square-wave reference signal and all initial conditions are set to zero.
	
	\begin{table}[t!]
		\caption{Gain and Delay Margins from the Loop Transfer Function for $\alpha$ Values.}
		\vspace{-0.25cm}
		\label{tbl1:margins}
		\includegraphics[width=\linewidth]{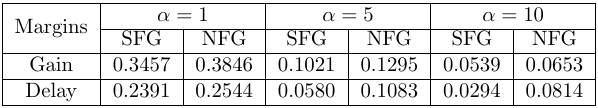}
		\vspace{-0.92cm}
	\end{table}
	
\begin{figure}[t!] 
		\hspace{0.33cm} \includegraphics[width=0.84\textwidth]{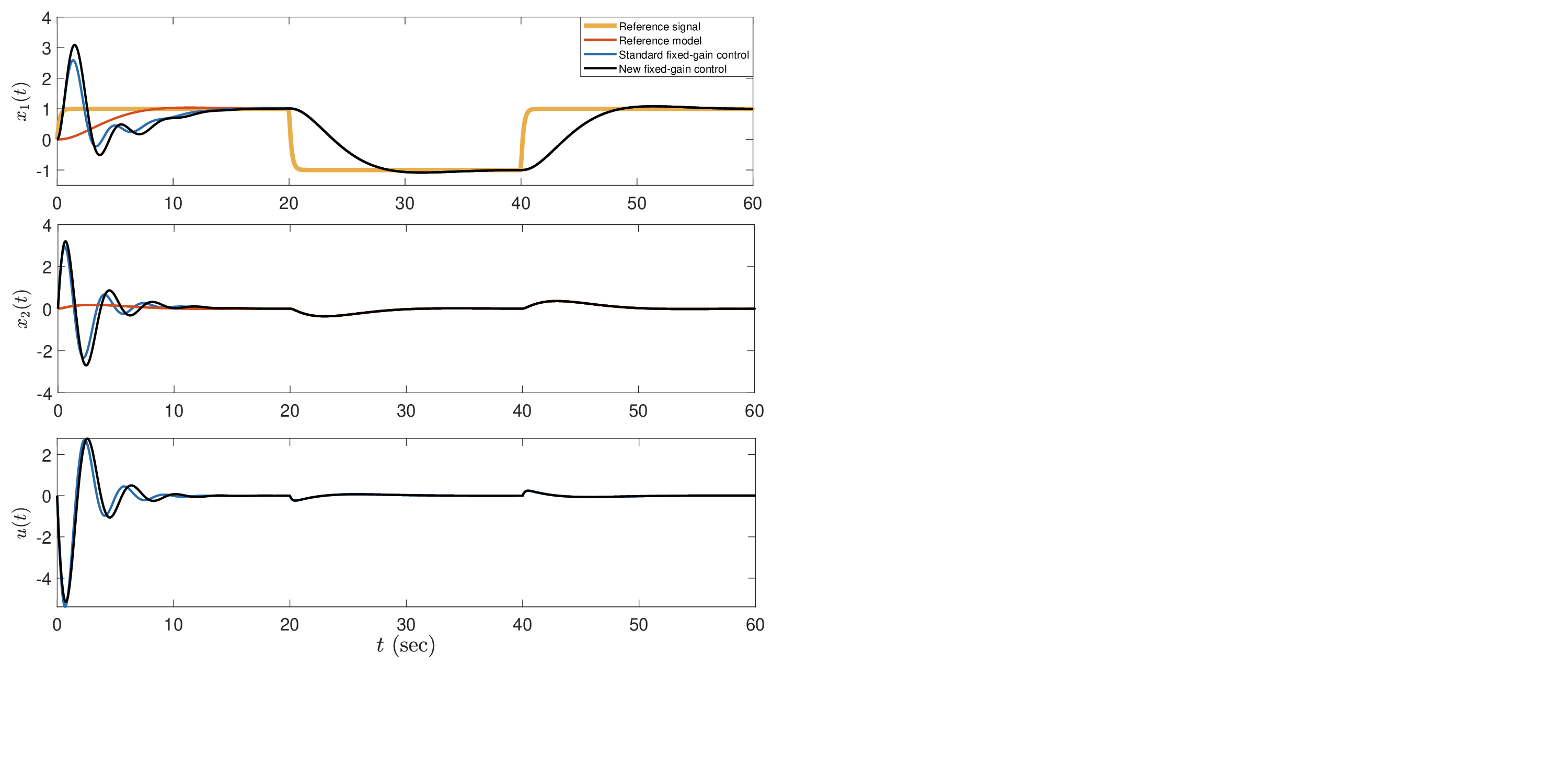} \vspace{-1.9cm}
		\caption{Closed-loop system performances with the symbiotic control framework having the standard and new fixed-gain controls when $\alpha=1$.} \vspace{-0.3cm}
		\label{Alpha_1} 
	\end{figure}

\begin{figure}[h!]
		\hspace{0.33cm}  \includegraphics[width=0.84\textwidth]{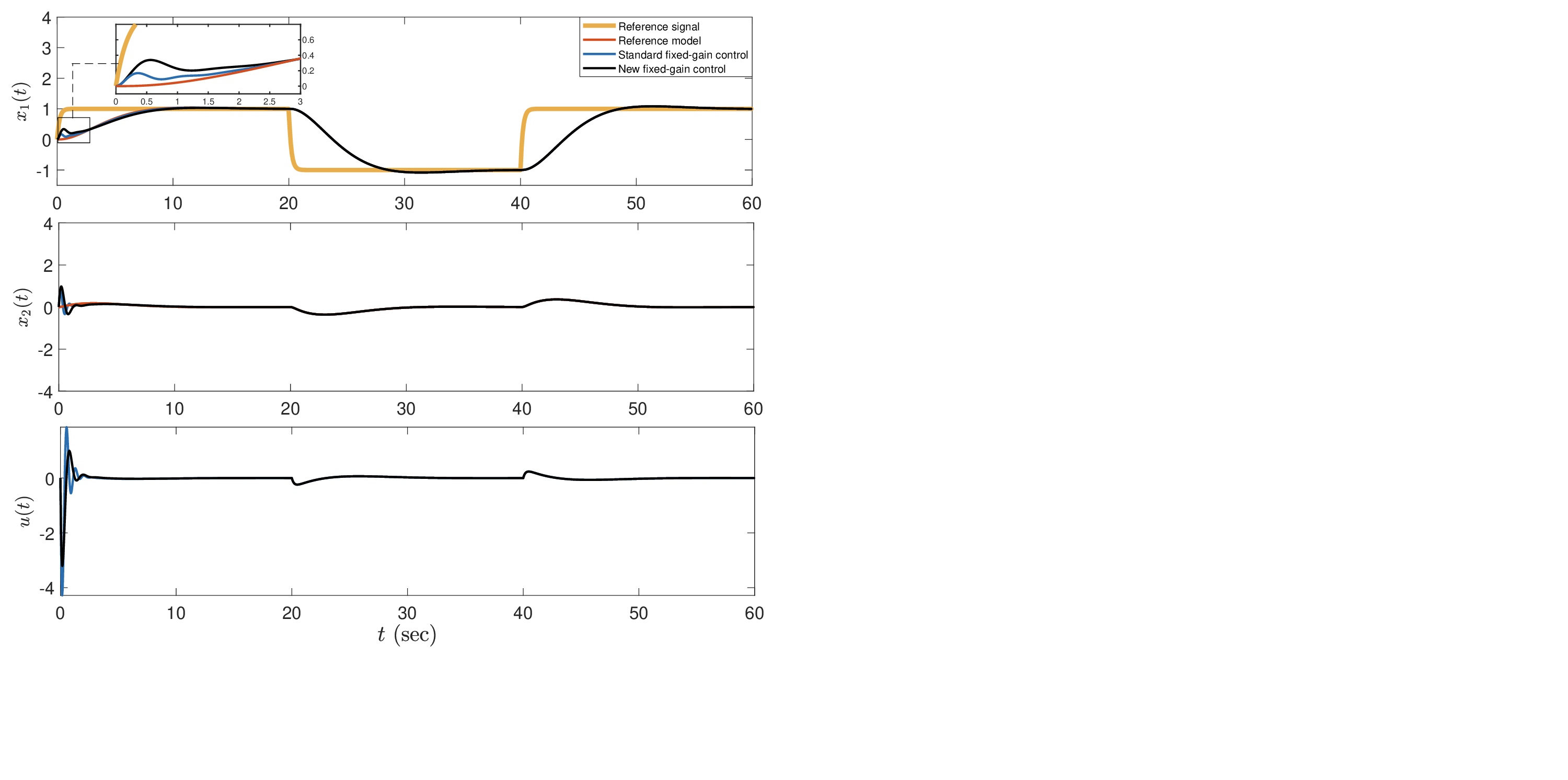}  \vspace{-1.9cm}
		\caption{Closed-loop system performances with the symbiotic control framework having the standard and new fixed-gain controls when $\alpha=5$.} \vspace{-0.3cm}
		\label{Alpha_5} 
	\end{figure}

\begin{figure}[h!]
		\hspace{0.33cm}  \includegraphics[width=0.84\textwidth]{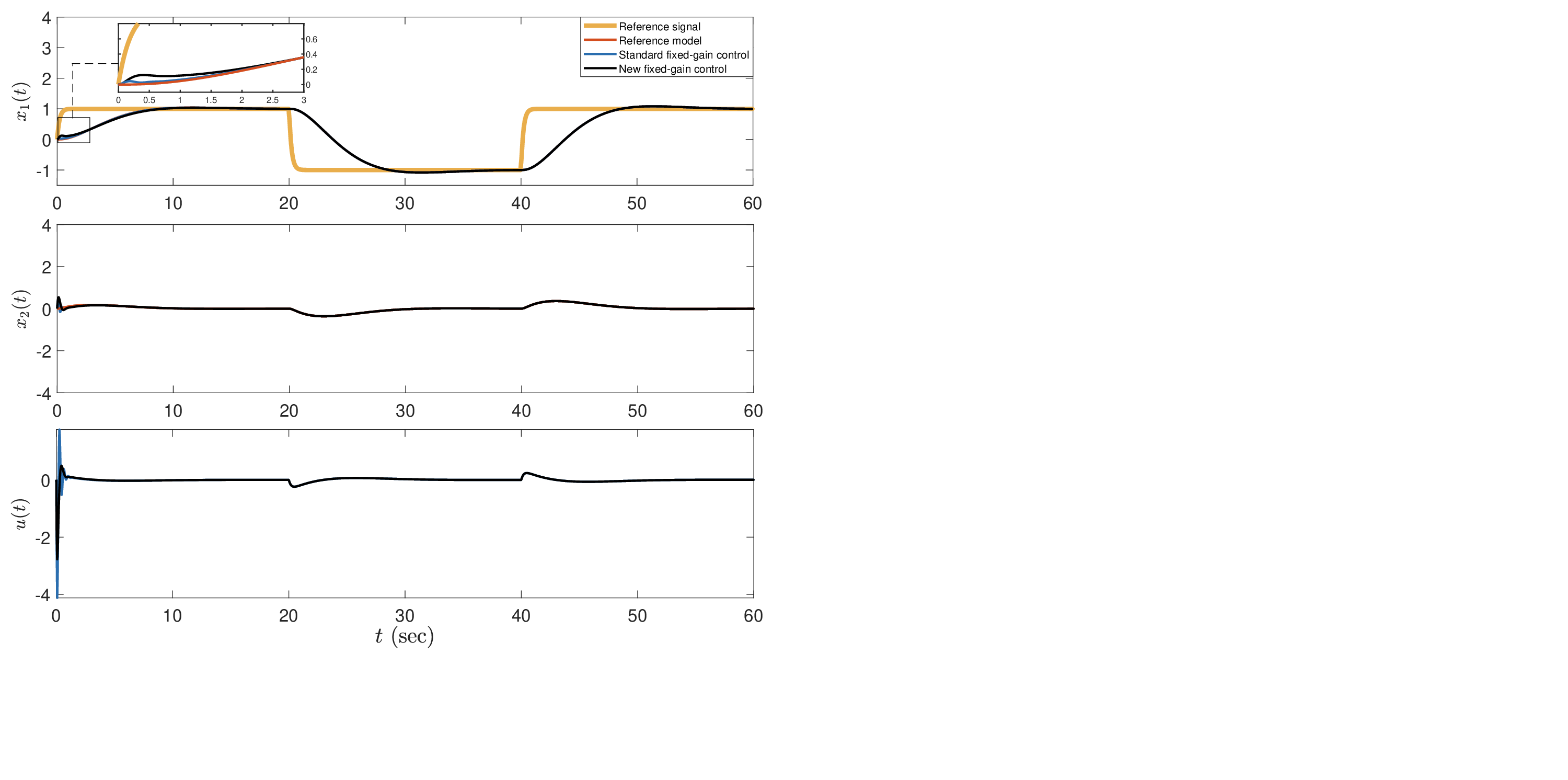} \vspace{-1.9cm}
		\caption{Closed-loop system performances with the symbiotic control framework having the standard and new fixed-gain controls when $\alpha=10$.} \vspace{-0.3cm}
		\label{Alpha_10} 
	\end{figure}

	In the first example, we select a range of fixed-gain control parameters (i.e., $\alpha$) for both the standard fixed-gain control and the new fixed-gain control to show the effect of this parameter on both stability margins and the performance of the dynamical systems.
	{For this example, we choose $\epsilon_1=3$ and $\epsilon_2=10$ to regulate the stability margins for the latter fixed-gain control method since this selection provides adequate stability margins without significantly compromising the closed-loop performance (see Table \ref{tbl2_1:margins}). }
	The responses of the closed-loop system with the standard fixed-gain control and the new fixed-gain control are shown in Figures \ref{Alpha_1}, \ref{Alpha_5}, and \ref{Alpha_10} when $\alpha=1$, $\alpha=5$, and $\alpha=10$, respectively.
	It is clear that the reference signal is perfectly tracked with both control methods after the transient period.
	Moreover, the transient performance of both control methods improves when we increase $\alpha$.
	This is consistent with the discussion in Remark 2. 
	Although increasing $\alpha$ helps the system behave as desired, this reduces the stability margins, which can be seen from Table \ref{tbl1:margins}. 
	This table presents gain and delay margins for the standard fixed-gain control (i.e., SFG) and the new fixed-gain control (i.e., NFG) when $\alpha=1$, $\alpha=5$, and $\alpha=10$.
	
	The performance of the closed-loop system with the standard fixed-gain control is slightly better than the new fixed-gain control for all $\alpha$ values. 
	Yet, the new fixed-gain control does not sacrifice the stability margins as much as the standard fixed-gain control when we increase $\alpha$.
	Specifically, the gain and delay margins for both control methods are close to each other when $\alpha=1$. 
	However, this changes when we increase $\alpha$ from $1$ to $10$.
	Especially, the delay margin of the new fixed-gain control (i.e., $0.0814$) is almost three times better than the delay margin of the standard fixed-gain control (i.e., $0.0294$) when $\alpha=10$.
	Similarly, the new fixed-gain control provides better gain margins for all $\alpha$ values. 
	
	
	
	\begin{figure}[t!] \vspace{-0.0cm}
		\hspace{-0.3cm}\includegraphics[width=1.33\textwidth]{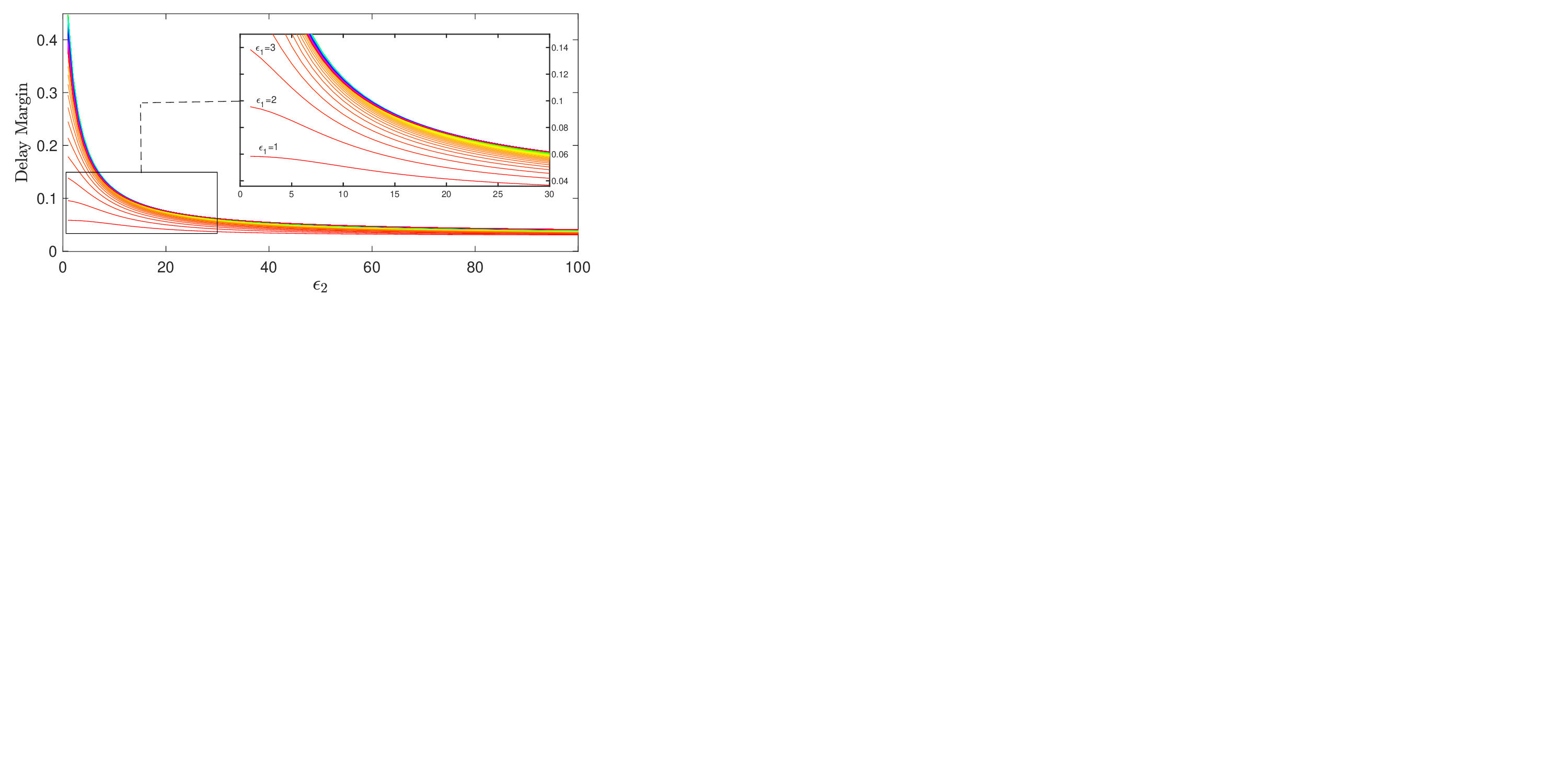}
		\vspace{-8cm}
		\caption{Delay margin based on $\epsilon_1$ and $\epsilon_2$ when $\alpha=10$.}
		\label{Delay_margin} 
		\hspace{-0.3cm}\includegraphics[width=1.33\textwidth]{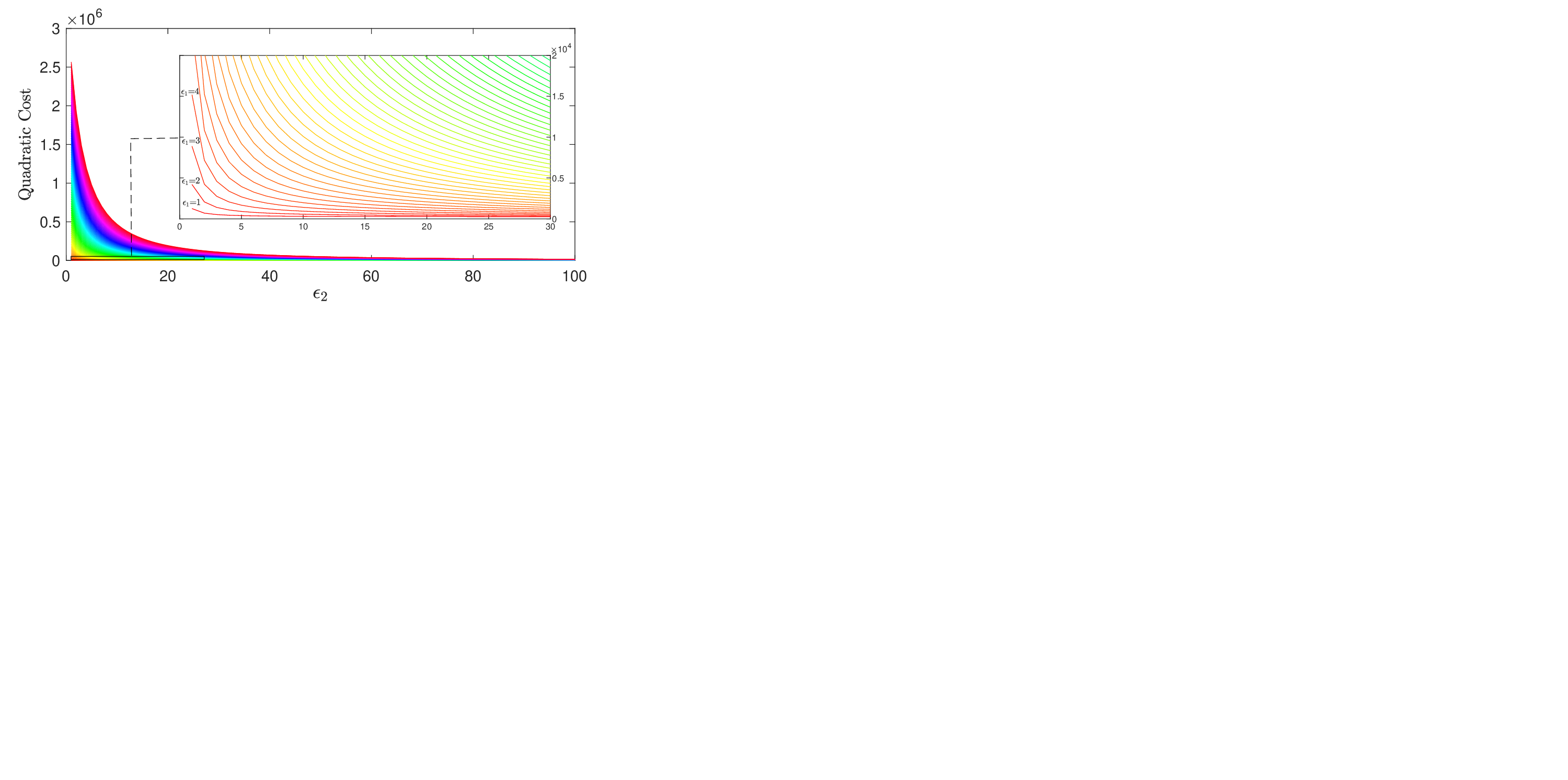} \vspace{-8cm}
		\caption{Quadratic cost based on $\epsilon_1$ and $\epsilon_2$ when $\alpha=10$.} 
		\vspace{-0.5cm}
		\label{Quadratic_cost}  
	\end{figure}
	
	Next, we demonstrate the importance of selection of the low-pass filter parameters (i.e., $\epsilon_1$ and $\epsilon_2$) on both the performance of the closed-loop system and the stability margins. 
	For this purpose, we define the following quadratic cost function $ p(e(t)) \hspace{-0.02cm}\triangleq\hspace{-0.02cm} \int_{0}^{t} e\mT(\tau)e(\tau)\rom{d}\tau$, which we use to compare the closed-loop performance based on the selection of $\epsilon_1$ and $\epsilon_2$. 
	In particular, Figure \ref{Delay_margin} is given to understand the effect of the low-pass filter parameters on the delay margin and Figure \ref{Quadratic_cost} is given to discuss the impact of the low-pass filter parameters on the closed-loop performance, where $\alpha=10$ is selected. 
	The following observations are now immediate\footnote{Note that solid lines in these figures represent a fixed $\epsilon_1$. 
		Note also that there is a 1-unit increment of $\epsilon_1$
		once solid lines move further away from $x\hspace{-0.06cm}-\hspace{-0.06cm}\rom{axis}$ as shown in the zoom part of these figures.}: 
	\begin{itemize}
		\item Increasing $\epsilon_1$ has a diminishing impact on the delay margin when $\epsilon_2$ is fixed. Beyond a certain point, it almost does not have an effect on the delay margin. However, it  deteriorates the performance of the closed-loop system. 
		\item If $\epsilon_2$ is fixed, then the performance of the closed-loop system improves with the selection of lower $\epsilon_1$.  
		\item Increasing $\epsilon_2$ has a diminishing effect on the performance of the closed-loop system when $\epsilon_1$ is fixed.
	\end{itemize}
	
	In the second example, we discuss the importance of selecting appropriate low-pass filter parameters to achieve better stability margins without compromising closed-loop performance, where we consider two scenarios. 
	In the first one, we aim for a quadratic cost of around $100$ in Figure \ref{Quadratic_cost}, which is satisfied by the selection of $\epsilon_1=3$ and $\epsilon_2=10$.
	We also obtain other combinations of $\epsilon_1$ and $\epsilon_2$ satisfying this cost\footnote{Note that the selection of these low-pass filter parameters provides the similar closed-loop performance with minor differences.
		Due to the page limitation, we do not provide the closed-loop performance of each selection of these low-pass filter parameters.}.
	The stability margins for these low-pass filter parameters are given in Table \ref{tbl2_1:margins}. 
	Although the closed-loop performance remains consistent, varying $\epsilon_1$ and $\epsilon_2$ can significantly improve stability margins as shown in Table \ref{tbl2_1:margins}.
	In the second scenario, we target a quadratic cost of $300$ in Figure \ref{Quadratic_cost}. 
	We again select $\epsilon_1$ and $\epsilon_2$ values satisfying this cost. 
	The stability margins for the selections of these low-pass filter parameters are summarized in Table \ref{tbl2_2:margins}. 
	It is evident that some $\epsilon_1$ and $\epsilon_2$ choices significantly enhance stability margins although there is no dramatic change in the closed-loop performance.
	
	\begin{table}[t!]
		\vspace{0.25cm}
		\caption{Gain and Delay Margins for example 2: scenario 1.}
		\vspace{-0.25cm}
		\label{tbl2_1:margins}
		\includegraphics[width=\linewidth]{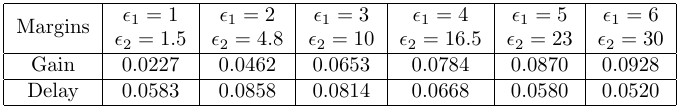}
		\vspace{-1cm}
		\vspace{0.3cm}
		\caption{Gain and Delay Margins for example 2: scenario 2.}
		\vspace{-0.25cm}
		\label{tbl2_2:margins}
		\includegraphics[width=\linewidth]{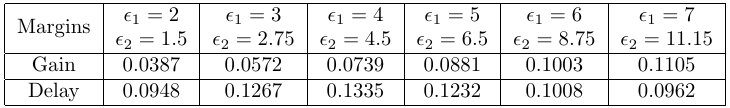}
		\vspace{-1cm}
	\end{table}

	
	\section{Conclusion}\label{conclusion}
	
	This paper presented a new fixed-gain control architecture incorporating a low-pass filter approach to address the stability margin regulation problem in the recently proposed symbiotic control framework. 
	Specifically, it was shown that this filter effectively limits the aggressive behavior of the fixed-gain control law, which in turn improves the gain and delay margins without significantly compromising closed-loop system performance. 
	Furthermore, the stability of the closed-loop system in the presence of exogenous disturbances was rigorously demonstrated in Theorems 1 and 2. 
	The new fixed-gain control architecture can be directly applied to mitigate the effects of parametric and nonparametric uncertainties within the recently proposed symbiotic control framework.
	{Future work will include optimization of $\alpha$, $\epsilon_1$, and $\epsilon_2$ since they play a critical role on the interplay between stability margins and closed-loop system performance. } 
	


	\bibliographystyle{IEEEtran} 
	\baselineskip 12pt
	\bibliography{refs}

\end{document}